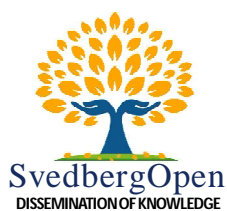

## International Journal of Management Research and Economics

Publisher's Home Page: https://www.svedbergopen.com/

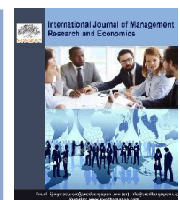

**Research Paper**  **Open Access**

# On the implementation of the universal basic income as a response to technological unemployment

Le Dong Hai Nguyen[1*]

[1]School of Foreign Service, Georgetown University, 3700 O St NW, Washington, DC 20057, USA. Email: ln406@georgetown.edu

**Abstract**

The effects of automation on our economy and society are more palpable than ever, with nearly half of jobs at risk of being fully executed by machines over the next decade or two. Policymakers and scholars alike have championed the Universal Basic Income (UBI) as a catch-all solution to this problem. This paper examines the shortcomings of UBI in addressing the automation-led large-scale displacement of labor by analyzing empirical data from previous UBI-comparable experiments and presenting theoretical projections that highlight disappointing impacts of UBI in the improvement of relevant living standards and employability metrics among pensioners. Finally, a recommendation shall be made for the retainment of existing means-tested welfare programs while bolstering funding and R&D for more up-to-date worker training schemes as a more effective solution to technological unemployment.

## 1. Introduction

SB Ekhad's *The Flamboyant Partridge* had been on the bestselling list last year. Eager to meet the writer at the Oxford Literary Festival, his fans were shocked to discover that their revered author was merely "a stack of computer hardware fronted by a screen that flickered on to reveal a human-like avatar face" (Sautoy, 2019). SB Ekhad was, in fact, 3B1, an AI machine created by the University of Oxford's Mathematical Institute. This incident embodies the automation that threatens to replace human jobs. With the cost of implementation shrinking and the robot-to-workers ratio skyrocketing, the effects of automation on our economy and society are more palpable than ever. According to Frey and Osborne (2017), 47% of jobs could be fully executed by machines "over the next decade or two." Across the globe, the risk posed by automation on the workforce is of great concern, with more severe impacts on manufacturing-focused developing countries like China, where 77% of jobs are forecasted to be replaced by automation (Frey *et al.,* 2016).

In response to the threat of mass displacement of labor due to automation, economists, politicians, and even the business community have come to see Universal Basic Income (UBI) as the panacea. UBI "experiments" have been implemented in Finland, Canada, California and even become official political platforms for most left-wing politicians from the UK's Green Party to US Presidential Candidate Andrew Yang, who argue that UBI would encourage unemployed people to find work, create savings, and boost economic growth. With the Covid-19 pandemic, there has been renewed interest in implementing UBI, even among the right side of the political spectrum. It is also endorsed by many tech giants, notably Mark Zuckerberg and Richard Branson, who hope UBI would "encourage creativity and innovation." However, the failure of similar programs—from the 1960s US Negative Tax Program to Finnish Basic Income Experiment—raises serious concerns about UBI not as a pragmatic but rather a harmful response to the situation. The goal of this paper is to assess the shortcomings of UBI vis-à-vis existing means-tested welfare programs and determine whether UBI is an effective policy in response to technological unemployment.

*\* Corresponding author: Le Dong Hai Nguyen, School of Foreign Service, Georgetown University, 3700 O St NW, Washington, DC 20057, USA. E-mail:  ln406@georgetown.edu*

## 2. Current and past basic income experiments

The recent Basic Income Experiment run by Kela, the Social Insurance Institution of Finland allocated €560 per month unconditionally to 5,000 unemployed citizens between the ages of 25 and 58 with the initial expectation of "promot[ing] active participation and giv[ing] people a stronger incentive to work" in response to the changing nature of work due to automation, (Kansaneläkelaitos, 2019). However, the two-year experiment ended in late 2018 with its preliminary results showing "no effects on employment." Compared to non-recipients, the participants receiving benefits from Kela showed little improvements, if not adverse outcomes. Specifically, as Table 1 shows, the total number of employment days for the treatment group was merely 0.39 longer, and self-employment incomes were actually €21 lower than those of the control group (Fabric *et al.*, 2019). This attests to the failure of one of the most prominent UBI programs.

| Table 1: Employment status and benefits paid out by Kela on average in 2017 | | | | |
|---|---|---|---|---|
| | **Treatment group** | **Control group** | **Difference** | *p*-value |
| Employment status | | | | |
| Days in employment (number of days) | 49.64 | 49.25 | 0.39 | 0.87 |
| Persons with earnings or income from self-employment (%) | 43.70 | 42.85 | 0.85 | |
| Earnings and income from self-employment, total (€) | 4,230 | 4,251 | -21 | |
| Benefits provided by Kela (€) | | | | |
| Unemployment benefits | 5,852 | 7,268 | -1415 | |
| Social assistance | 941 | 1,344 | -403 | |
| Housing allowance | 2,525 | 2,509 | 16 | |
| Sickness allowance | 121 | 216 | -96 | |
| Number of observations | 2,000 | 173222 | | |
| Note: (i) The days in employment are based on data on accrual periods from the Finnish Centre for Pensions. Days of employment are defined as periods in the open labour market for which the calculated daily wage amounts to at least 23.74 euros; (ii) The percentage who have received earnings or income from self-employment and their number is based on data from the Finnish Tax Administration; (iii) The data on benefits provided by Kela is based on data from Kela's benefit register; (iv) The p-value shows the level of significance at which the equality of the averages for the treatment group and the control group can be cancelled out. Typically the difference between the groups is considered statistically significant when the p-value is 0.05 or smaller. | | | | |

From a historical standpoint, UBI or comparable policies dis-incentivized the unemployed population from seeking work. For example, from 1968 to 1980, the United States introduced four experiments of Negative Income Tax (NIT) on thousands of people in six states. NIT allowed people earning below a designated threshold to receive a supplemental paycheck from the government instead of paying taxes. In a way, NIT is a tax-deductible, exclusive UBI for the poor. However, the results of NIT experiments in the US show that compared to those not in the program, desired working hours decreased between 5% to 7.9% for men and between 17% to 21.1% for married women with children (Wiederspan *et al.*, 2015). Noticeably, the incentive to work "among single men, as claimed by West (1980), fell some 43% below non-recipients." The failure of the NIT program—which was analogous to UBI in various aspects—raises serious concerns about the beneficiaries' incentive to work and the resulting long-term consequences for the economy. Since UBI derives its resources from taxpayers' money, when the motivation to work declines, those who do work will have to pay the bill for those who do not, which in turn encourages more people not to work, thereby stifling the possibility of "growth, innovation and entrepreneurship" as claimed by UBI advocates such as Widerquist *et al.* (2013). In the long run, this will cause lower economic input and lower tax revenue to invest in the future, not to mention the unavoidable social tensions. Furthermore, providing basic incomes to the lowest earners could greatly dis-incentivize the least skilled and the most vulnerable victims to work, leaving less taxable income for the government to levy for other public functions.

## 3. Financing a basic income program

In addition to proving ineffective in the mitigation of unemployment, UBI comes at an exorbitant cost to taxpayers. Finnish failed *Basic Income Experiment* on only 5,000 people costs €20m in two years, according to its website, while Andrew Yang's proposal UBI plan would require a hefty $2.8 tn every year (234 million American citizens above 18 years of age, according to Howden and Meyer (2010) to be provided with $12,000 annually), which would be more than half of the current $4.4 tn US budget in 2019. With an already massive budget deficit, the cost of UBI would translate into a greater burden borne by American taxpayers' or further increase its national debt. We estimated that to grant a small sum of $1000 a month, most countries in the developed world would have to allocate from 35% to 50% of their GDP. Because of the massive funding required, UBI advocates suggest higher corporate taxes as well as new taxes on companies' market capitalization, including IPOs and mergers. However, a study among OECD countries, as shown in Figure 1, implies that "[GDP(PPP)] declines by 1.3% for each 10% points increase in the [corporate] tax rate" (Kopits, 2017). Therefore, implementing a higher corporate tax to fund UBI is not economically efficient.

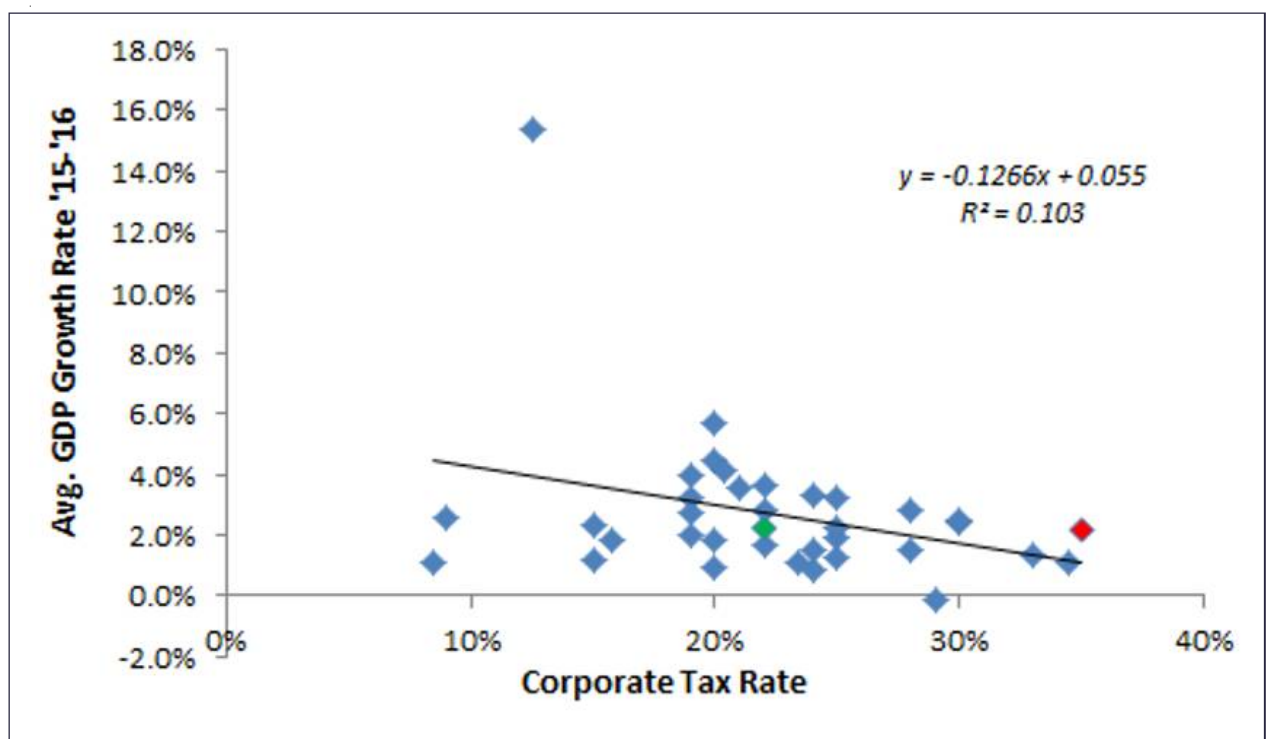

**Figure 1: Growth vs. Corp tax rate – All OECD countries**

Facing the difficulty of levying more tax for UBI, many countries especially developing ones, would choose to eliminate all existing means-tested welfare state programs and substitute them using a uniform UBI-styled grant to all citizens, such as the Indian Government's recent proposal. If governments chose this approach, the consequences would be catastrophic. Reed and Lansley (2016) claim that handing out $392 (£292) monthly to every adult while eradicating existing means-tested programs would cause "child poverty to increase by 10%, poverty among pensioners by 4%, and poverty among the working population by 3%." A potential explanation for this is, unlike means-tested welfare programs, UBI grants a fixed amount of money to all citizens uniformly, regardless of economic status. Since UBI also grants money to the upper and middle-classes, whose marginal benefit from it is minimal compared to disadvantaged ones who need welfare assistance the most, this approach is therefore irresponsible and unfair as money could have

been targeted at needy people at a much greater quantity in the case of existing means-tested programs. Even if UBI is designed as an "add-on" and all means-tested programs are to remain, which would add a massive figure to the budget, the results are still quite disappointing "with a modest effect on poverty:" "[f]or working-age people[, poverty] decreases less than 2 points (13.9% to 12%), and among pensioners it declines only 1 point (14.9% to 14.1%)" (Reed and Lansley, 2016).

## 4. Finding alternative solutions

From these examples, we can see that the massive expenditure on UBI is neither economically efficient nor socially equitable. More importantly, it does not address the problem of large-scale displacement due to automation. With nearly half of all jobs at risk of being automated over the next 20 years, it is in urgent need to establish systematic transition of workers from low and mid-level into high-level skilled professions in developing and emerging industries like AI, robotics and blockchain, as well as non-tech sectors that are not easily replicated by machines. As the old saying goes, "it is better to teach a man how to fish than to give him a bucket of fish. This requires an "automation-proof" education that truly prepares and adapts the labor force for fundamental changes in the workplace. As automation is part of the Fourth Industrial Revolution and as a similar trend with the previous three revolutions, the number of new jobs created would outweigh the ones displaced by machines. In the 1980s, when the personal computer was introduced, concerns regarding its threat of replacing people's work arose. Until 2015, the introduction of the PC has indeed displaced 3.5 million jobs in the US— but at the same time, it has created 19.2 million new ones (Bughin *et al.*, 2017). Therefore, the government's education policies to ensure the future workforce's transition into new emerging positions are critically crucial. This could be done via these approaches:

Firstly, the government needs to provide more vocational training that will enable displaced workers to enter high-growth sectors like renewable energy, which is expected to add 20 million new jobs by 2030, or the healthcare industry, where 80-130 million new opportunities are available by 2030 (Renner *et al.*, 2008).

Secondly, as low-skilled jobs are replaced by robots, more jobs will shift to high-level industries: "80% of the new jobs created will require a graduate-level education or equivalent" (Talwar *et al.*, 2018). Therefore, the government needs to encourage employees to continue higher education while still in employment to adapt to the higher skill level demanded. As historically proven, higher educational attainment is linked with lower levels of unemployment, for low skilled workers are more susceptible to becoming unemployed during economic downturns, as shown in the data from the Labor Department in Figure 2.

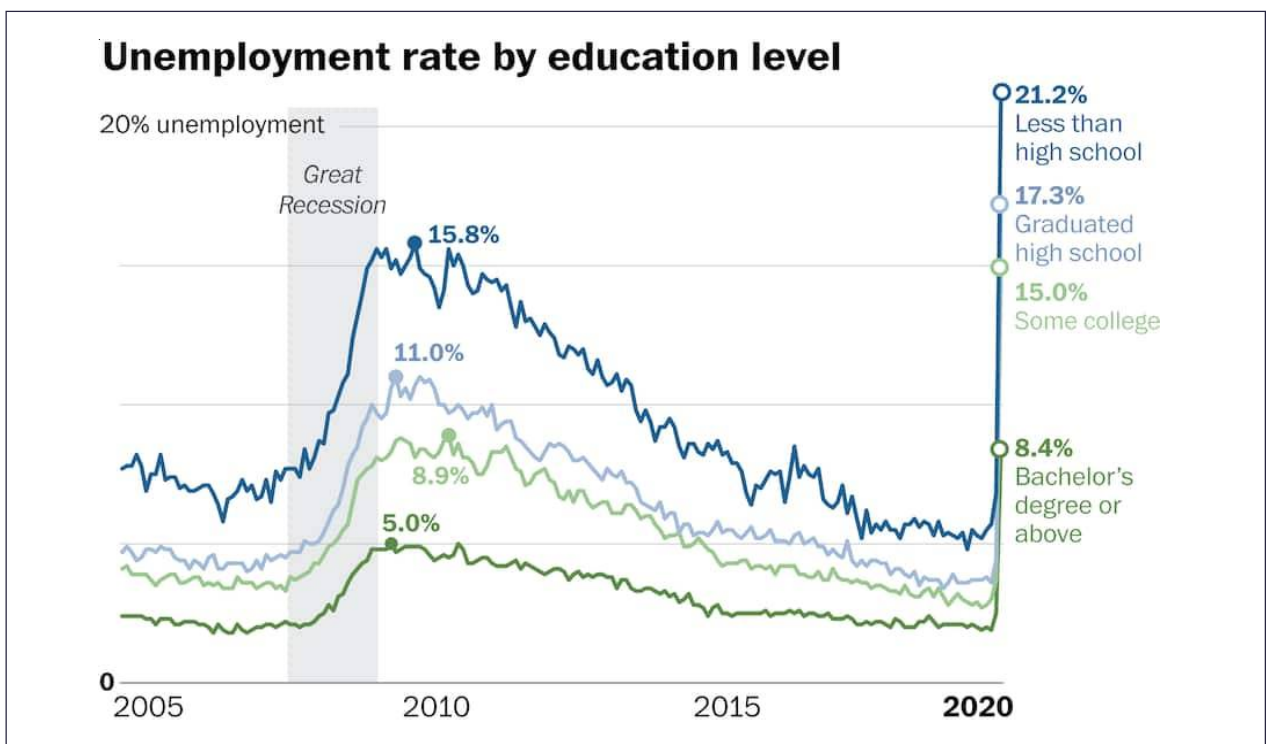

**Figure 2: Unemployment rate by educational level**

Thirdly, schools will also need to foster critical thinking, creativity, emotional intelligence, flexibility, and cultural agility—skills that cannot be replicated by machines. This would require an overhaul of our education system, demanding considerable funding from the government. Education, after all, is a much more affordable investment compared to UBI. The World Bank's World Development Indicators (2010), as shown in Figure 3, estimates that for every dollar the government spends on education, GDP grows on average by $20.

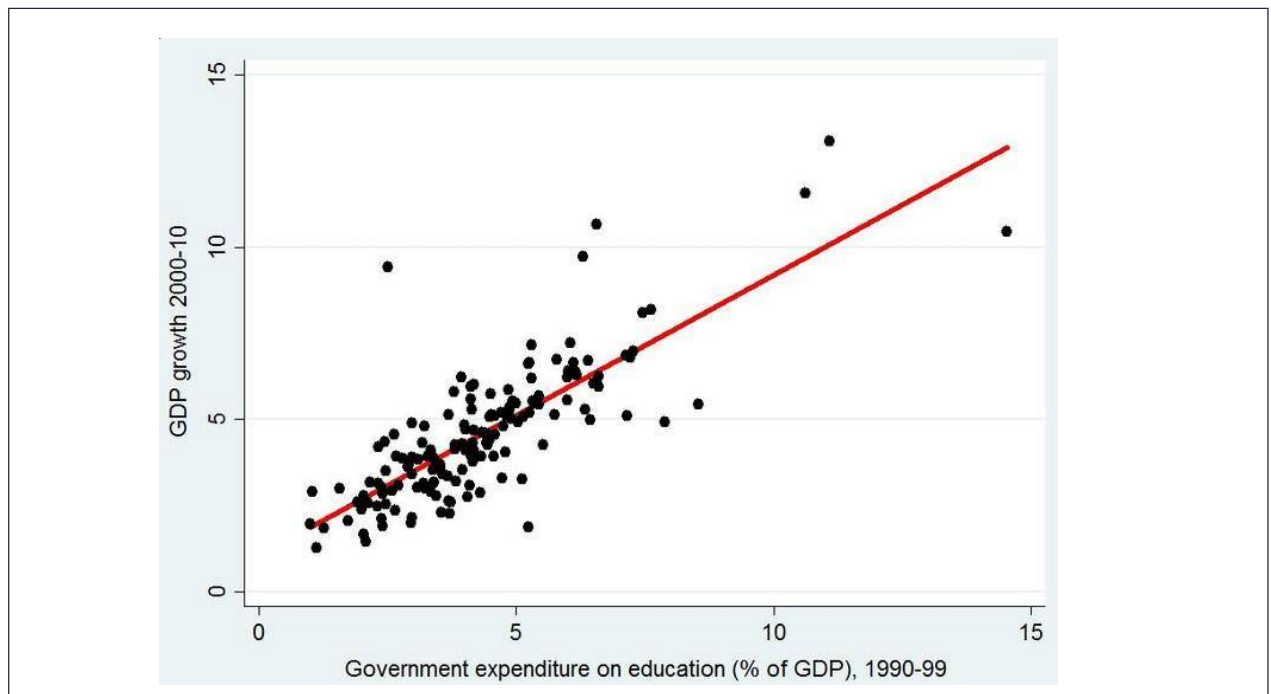

**Figure 3: Time-lagged chart showing the relationship between GDP measured by its annual growth rate between 2000 and 2010 and education expenditure measured as a share of total GDP over the period 1990-1999**

## 6. Conclusion

Results from two prominent UBI-analogous experiments show that granting a sum of money unconditionally to a population does not have a substantial nor positive effects on the employability, income, and other measures of living standards. Further investigations into the costs of implementation of UBI in addition to existing welfare programs suggest an exorbitant cost that would significantly enlarge a country's budget while the marginal benefit is minimal. In the hypothesis of replacing the welfare state with a UBI, living standard metrics such as poverty rate worsen substantially. We then consider and make a case for the alternative policy of retaining existing welfare programs with an emphasis on increasing funding for the education system and training initiatives for displaced and low-skilled workers.

### Acknowledgment

An early version of this manuscript won a high commendation from the Royal Economic Society and was featured among the *Young Economist of the Year* honorees in late 2019.